\documentclass[11pt]{iopart}
\usepackage{iopams}
\usepackage{epsfig}
\newcommand{\be}{\begin{equation}}
\newcommand{\ee}{\end{equation}}

\newcommand{\BE}{\begin{eqnarray}}
\newcommand{\EE}{\end{eqnarray}}

\newcommand{\sgn}{{\rm sgn}}
\newcommand{\erf}{{\rm erf}}
\newcommand{\id}{{\rm 1\!\!I}}

\newcommand{\bq}{\ensuremath{\mathbf{q}}}

\newcommand{\bR}{\ensuremath{\mathbf{R}}}

\newcommand{\avg}[1]{\left\langle{#1}\right\rangle}
\newcommand{\davg}[1]{\left\langle\left\langle{#1}\right\rangle\right\rangle}

\begin{document}
\title[Adapting to heterogeneous comfort levels]{Adapting to 
heterogeneous comfort levels}

\author{Luca De Sanctis\dag~ and Tobias Galla\dag\ddag
}

\address{\dag\ The Abdus Salam International Center for 
Theoretical Physics, Strada Costiera 11, 34014 Trieste, Italy}
\address{\ddag INFM-CNR, Trieste-SISSA Unit, V. Beirut 2-4, 
34014 Trieste, Italy}

\begin{abstract}
We study the learning dynamics of agents who adapt to heterogeneous
comfort levels in the context of an El-Farol type game, and show that
even an infinitesimal degree of heterogeneity in the resource levels
leads to a significant reduction of the fluctuations of the collective
action, and removes the phase transition observed in models with
homogeneous comfort level.  Our analysis is based on dynamical methods
of disordered systems theory, in particular on a generating functional
approach, and confirmed by numerical experiments. We also report on
simulations of a system in which the comfort levels fluctuate in time,
and point out crucial differences between models in which the comfort
levels of the agents fluctuate collectively and individually
respectively.  Finally we comment on a possible characterisation of
El-Farol and Minority Games according to the presence or absence of
ergodicity-breaking phase transitions at infinite integrated response.
\end{abstract}

\pacs{02.50.Le, 87.23.Ge, 05.70.Ln, 64.60.Ht}

\ead{\tt lde\_sanc@ictp.it, galla@ictp.it}

\section{Introduction}
Complex adaptive systems of inductive agents often display a remarkably
rich global behaviour, which cannot be explained straightforwardly from
their microscopic interactions. The tools of statistical mechanics
have here been seen to be able to provide further insight, and allow
for theoretical progress for a variety of agent-based models. Most
notably, analytical solutions for the Minority Game (MG), a
mathematical abstraction of the El-Farol bar problem, have been
obtained using equilibrium and non-equilibrium techniques of
disordered systems theory \cite{Book1,Book2,Book3}.

In the El-Farol bar problem \cite{arthur}, $N=100$ customers have to
decide independently whether or not to attend a concert in the bar,
the latter having a capacity of $\lambda=60$ seats. Agents will in
general not enjoy the bar if it is too crowded, i.e. if more than
$\lambda$ agents attend. Thus depending on the attendance $A$, agents
who go to the bar are considered winners if $A<\lambda$, and
conversely, agents who decide to stay home win if $A>\lambda$. We will
refer to $\lambda$ as the {\em comfort level} in the
following. Players in the El-Farol bar problem are inductive agents:
they use individual `predictors', based on the past attendance, to predict
whether or not the bar will be crowded at the next time step, and
according to these predictions they then decide whether to attend or
not. They learn by experience and employ the most accurate predictor which they can access.

Under fairly weak assumptions it can be shown that the temporal
average of the attendance, $\avg{A}$, converges to the comfort level
$\lambda$ in Arthur's El-Farol model provided the predictors are not
systematically biased \cite{arthur}. A second question relates to the degree to
which agents are able to reduce the fluctuations of $A$ about 
$\lambda$, i.e. to the quantity $\sigma^2=\avg{A^2}-\avg{A}^2$, where the brackets denote an average over time. We will refer to $\sigma^2$ as the {\em volatility} in the following.

The MG in its original form \cite{ChalletZhang} is a mathematical
formulation of the El-Farol problem at comfort level
$\lambda=N/2$. All agents take binary decisions at each time-step, and
players in the minority group win. Predictors, fixed at the beginning
of the game and randomly generated, are here unbiased in the sense
that they advise to attend or stay home with equal statistical
weights. Generalisations to $\lambda\neq N/2$ and/or systematically
biased strategies are possible, and have been studied by numerical
simulations in
\cite{johnson1,johnson2}. More systematic studies of El-Farol
games with general {\em uniform} comfort levels have been presented
\cite{ChalMarsOtti03}, see also \cite{damien}.

In this paper we consider an El-Farol type problem with {\em
heterogeneous} comfort levels, so that each agent $i$ prefers to
attend the bar only if $A<\lambda_i$, where $\lambda_i$ varies across
the population of agents. Our work is an extension of
\cite{ChalMarsOtti03}, and we reproduce some results reported there as
a special case. We here employ a dynamical approach whereas the
analysis of \cite{ChalMarsOtti03} is based on static replica calculations. Both approaches are now standard in the context of MGs and related models \cite{Book1,Book2}.

The purpose of this study is here twofold. Firstly we aim at
understanding the role of heterogeneous comfort levels for the
learning dynamics of the agents. In particular we will be concerned
with the question of whether or not heterogeneity in the comfort
levels compromises the ability of the population of agents to converge
to a `mean' comfort level. Similarly we will study the influence of
heterogeneity in the comfort levels on the temporal fluctuations of
the attendance. The second reason for studying the present system
rests in more theoretical issues related to the phase behaviour of MGs
and related models. In the standard MG a phase transition between an
ergodic and a non-ergodic regime has been found, and identified with a
point in parameter space in which the integral over the response
function of the system becomes singular \cite{Book2,Book3}. In the
non-ergodic regime a continuum of attractors of the learning dynamics
appears to be present and introduces interesting effects such as a
marked sensitivity to initial conditions, which is absent in the
ergodic phase.

Some variants of the MG, however, which are typically only slight
modifications of the original game have been seen not to display the above transition. Some variants exhibit no transition at all, others one that is of a different type than the one described above. The transition is absent for
example in so-called grand-canonical MGs (GCMG) \cite{gcmg}, and
similarly such an absence may be suspected for MG models with finite
score memories \cite{mem} although no analytical results are available
as yet. In MGs with impact correction \cite{impact} and dilution of
the agents' interaction matrix \cite{dilute} the transition of the
type as in the original MG is preceded by one which is referred to as
`memory-onset' transition \cite{Book2}. Ergodicity breaking here still
occurs, but at finite integrated response. This type of transition has
been associated with replica-symmetry breaking in static studies. The second purpose of this paper is thus to try to shed some
more light on the circumstances under which MG models display a
transition of the type of the original MG. We here show that MGs with
any degree of heterogeneity in the comfort levels do not display this
type of transition, and conclude with some conjectures on how the
structure of the agents' learning dynamics may determine the presence
or otherwise of such a transition.

\section{Definition of the model} 

We consider a system of $N$ agents, labelled with Roman indices
$i,j\in\{1,\dots,N\}$. At each round $t$ of the game each agent $i$
takes a binary decision $b_i(t)\in\{-1,1\}$ in response to the
observation of a common piece of publicly available information,
labelled by $\mu(t)$. $b_i(t)=1$ may here correspond to player $i$
attending the bar at time $t$, and $b_i(t)=-1$ to him not
attending. While the information encodes the actual history
(i.e. the previous attendances) in the original version of the
El-Farol problem and in the original MG, we will here assume that
$\mu(t)$ is chosen randomly and independently from a set with
$P=\alpha N$ possible values at any $t$,
i.e. $\mu(t)\in\{1,\dots,\alpha N\}$. This has been seen not to alter the
qualitative behaviour of different variants of the MG
\cite{Book1,Book2}. $\alpha=P/N$ is here the main control parameter of the model, and taken not to scale with $N$. One then defines the re-scaled total outcome at
round $t$ as $A(t)=N^{-1/2}\sum_i b_i(t)$, resulting in $-\sqrt{N}\leq
A(t)\leq \sqrt{N}$ (in steps of $2/\sqrt{N}$). Note that in this
notation ($b_i(t)\in\{-1,1\}$ as opposed to $b_i(t)\in\{0,1\}$ for
not-attending/attending the bar) an attendance of $N/2$ in the
original El-Farol problem corresponds to $A=0$, and accordingly the
comfort level in an actual MG (in which the bar is considered crowded
if more than precisely half of the agents attends) would be
$\lambda=0$. Agents who prefer a bar filled by less than half have
$\lambda_i<0$, agents who can enjoy the bar even if it is populated by
more than $N/2$ attendees have $\lambda_i>0$. In general the
attendance in the original El-Farol problem can be obtained as
$(\sqrt{N}A(t)+N)/2$ from the global action $A(t)$, in our
conventions. Having this simple linear transformation in mind, we will
refer to $A(t)$ as the attendance at time $t$ in the following, and,
as mentioned above, to $\lambda_i$ as the comfort level of player
$i$. Note that $A(t)$ remains a well-defined finite quantity in the
thermodynamic limit $N\to\infty$ at $\alpha=P/N$ finite, with which
the statistical mechanics analysis of the model is concerned.

In order to take their decisions on whether or not to attend each
agent $i$ holds two fixed trading strategies (look-up tables)
$\bR_{i,a}=(R_{i,a}^1,\dots,R_{i,a}^P)\in\{-1,1\}^P$, with $a=\pm 1$.
The binary entries $R_{i,a}^\mu$ are drawn from some distribution (to
be specified below) before the start of the game; these entries (along
with the comfort levels $\lambda_i$) represent the quenched disorder
of this problem.  If agent $i$ decides to use strategy $a$ in round
$t$ of the game, his action at this stage will be
$b_i(t)=R_{i,a}^{\mu(t)}$. The agents decide which strategy to use
based on points $p_{i,a}(t)$ which they allocate to each of their
strategies. These virtual scores are based on the player's potential
success had he always played that particular strategy, and are
updated as follows at every time-step:
\begin{equation}
p_{i,a}(t+1) = p_{i,a}(t) - R_{i,a}^{\mu(t)} (A(t)-\lambda_i). \label{eq:scores}
\end{equation}
Strategies which
would have produced a decision to the liking of player $i$ are thus
rewarded, i.e. if $A(t)>\lambda_i$ the scores of strategies $a$ with
$R_{i,a}^{\mu(t)}<0$ are increased and vice versa for
$A(t)<\lambda_i$. In the MG literature the above type of process
(\ref{eq:scores}) with an explicit dependence on the influx of
information patterns $\mu(t)$ is generally referred to as `on-line
dynamics'. At each round $t$ each player $i$ then uses the strategy in
his arsenal with the highest score, i.e. $b_i(t)=R_{i,s_i(t)}^{\mu(t)}$, where $s_i(t) =
\mbox{arg max}_a\, p_{i,a}(t)$. It remains to specify the statistics of
the $\{R_{i,a}^\mu\}$. We here choose
\be
P(R_{i,a}^\mu)=\frac{1+\rho_i/\sqrt{N}}{2}\delta_{R_{i,a}^\mu,1}
+\frac{1-\rho_i/\sqrt{N}}{2}\delta_{R_{i,a}^\mu,-1},
\ee
as the distribution from which the $R_{i,a}^\mu$ are drawn (with
$\delta_{x,y}=1$ if $x=y$ and $\delta_{x,y}=0$ otherwise), so that
$\overline{R_{i,a}^\mu}=\rho_i/\sqrt{N}$ for all $a,\mu$
($\overline{\cdots}$ here denotes an average over the quenched
disorder). In the limit of large $N$, which we will eventually
consider, the distribution of the $R_{i,a}^\mu$ is well-defined for any
$\rho_i\in\mathbb{R}$. If $\rho_i>0$ then player $i$'s strategies are
more likely to advise him to take a positive action ($b_i=+1$) than a
negative one, and vice versa. We will refer to $i$'s strategies as
being {\em consistent} with his comfort level $\lambda_i$ if
$\rho_i=\lambda_i$, and as {\em inconsistent} otherwise. For an agent
with consistent strategies the probability of him attending the bar
upon randomly choosing one of his strategies (regardless of their
scores) corresponds to his comfort level. The scaling of the comfort
levels and strategy biases with $N$ has here been chosen to ensure a
well-defined thermodynamic limit, and corresponds to the choices of
\cite{ChalMarsOtti03}.

In the following we will consider a population of agents in which all
$\rho_i$ are drawn independently from a fixed distribution $Q(\rho)$,
and the $\lambda_i$ from $R(\lambda)$.  The standard MG is recovered
for $\lambda_i=\rho_i\equiv 0~\forall i$, i.e. $Q(\rho)=\delta(\rho)$
and $R(\lambda)=\delta(\lambda)$.  As we will see in the further
analysis, the specific details of the distribution $Q(\rho)$ are
unimportant, and the only feature relevant for the properties of the
model is the mean strategy bias $\rho_0=\int d\rho Q(\rho)\rho$. We
will also demonstrate that the model is invariant against
simultaneous uniform shifts of the $\rho_i$ and $\lambda_i$,
i.e. under $\lambda_i\to\lambda_i+\Delta$ and $\rho_i\to\rho_i+\Delta$
for any $\Delta\in\mathbb{R}$. Without loss of generality we will
therefore mostly restrict to cases with $\rho_0=0$ in the following.

Finally the key observables we will study in this model are given by
the deviation of the mean attendance from the mean comfort level
$\lambda_0=\int d\lambda R(\lambda)\lambda$, i.e.  by
$|\avg{A}-\lambda_0|$, and by the fluctuations of the attendance about
its mean, i.e. by $\sigma^2=\avg{A^2}-\avg{A}^2$. $\avg{\cdots}$ here
stands for an average over time in potential stationary states of the
system, i.e.  after some sufficiently long equilibration time.  To
conclude this section we note that that the model of
\cite{ChalMarsOtti03} can be recovered upon assuming homogeneous
comfort levels and strategy biases, i.e. by setting
$\rho_i\equiv\rho_0$ and $\lambda_i\equiv\lambda_0$ for all $i$.
\section{Statistical mechanics analysis}
We will here pursue a dynamical approach based on a generating
functional analysis of the score update rules. We here consider the
so-called batch process of the learning dynamics, which for the present case
reads \BE\label{eq:batch}
q_i(t+1)=q_i(t)-\frac{2}{\sqrt{N}}\sum_{\mu}\xi_i^{\mu}\left[
\frac{1}{\sqrt{N}}\sum_j\left\{ \xi_j^\mu
s_j(t)+\omega_j^\mu\right\}-\lambda_i\right]+h_{\lambda_i}(t),
\EE similar to that of the standard MG \cite{HeimCool01,Book2}. We have here
introduced the score differences
$q_i(t)=\frac{1}{2}\left(p_{i,1}(t)-p_{i,-1}(t)\right)$ and the
quantities
$\xi_i^{\mu}=\frac{1}{2}\left(R_{i,1}^{\mu}-R_{i,-1}^{\mu}\right)$ and
$\omega_i^{\mu}=\frac{1}{2}\left(R_{i,1}^{\mu}+R_{i,-1}^{\mu}\right)$,
using conventions which are now standard in the MG-literature.
$s_j(t)\in\{-1,+1\}$ indicates the strategy player $j$ is using at
time $t$ and is given by $s_j(t)=\sgn[q_j(t)]$, so that player $j$'s
action at time $t$ reads $\omega_j^{\mu}+\xi_j^{\mu}s_j(t)$ on the
occurrence of information pattern $\mu$. $h_{\lambda_i}(t)$ is an
external perturbation field, which is set to zero in all simulations
and mostly a mathematical device added to generate response functions.
Perturbations are assumed to be identical for agents with identical
comfort level, hence the subscript $\lambda_i$. Note also that
compared to the on-line process an effective average over all
information patterns $\mu$ has been performed at each time-step and
that time has been re-scaled. 

The above batch process is the starting point for the dynamical
analysis of the problem based on generating functionals. This
technique is now standard in the context of the MG and we will only
report the final outcome of the theory here, with some additional
information in the appendix. For the further mathematical details of
the computation in similar cases we refer 
to the recent textbook \cite{Book2}.

The generating functional analysis turns the Markovian problem of
interacting agents into a self-consistent description in terms of
decoupled effective agents subject to non-Markovian stochastic
processes, and is exact in the thermodynamic limit $N\to\infty$ (at fixed $\alpha=P/N$). For an effective agent with strategy bias $\lambda$ this
process reads 
\BE q_\lambda(t+1)=q_\lambda(t)-\alpha\sum_{t'\leq t}
(\id+G)^{-1}_{tt'}s_\lambda(t')+\sqrt{\alpha}\eta_\lambda(t)+h_\lambda(t),
\EE 
where the second term on the right-hand side marks a retarded
interaction in time and renders the process non-Markovian.
$s_\lambda(t)$ is given by $s_{\lambda}(t)=\sgn[q_\lambda(t)]$ and
$\eta_{\lambda}(t)$ represents the stochasticity of this process. More
precisely, $\eta_\lambda(t)$ is Gaussian noise of zero mean, and with
temporal correlations 
\BE
\hspace{-2cm}\avg{\eta_\lambda(t)
\eta_{\lambda}(t')}_\star=[(\id+G)^{-1}D(\id+G^T)^{-1}]_{tt'}+2f_t
f_{t'}-2\lambda(f_t+f_{t'})+2\lambda^2E_{tt'}.  \EE Here
$\id_{tt'}=\delta_{tt'}$ is the identity matrix and $E_{tt'}=1~\forall
t,t'$, and $\avg{\cdots}_\star$ refers to an average over realisations
of the effective process (i.e. over realisations of the
$\eta_\lambda$). $C$ and $G$ are the correlation and response
functions of the system respectively 
\BE
C_{tt'}=\lim_{N\to\infty}N^{-1}\sum_i \overline{\davg{s_i(t)s_i(t')}},
~~~~ G_{tt'}=\lim_{N\to\infty}N^{-1}\sum_i
\frac{\partial\overline{\davg{s_i(t)}}}{\partial h_{\lambda_i}(t')}
\EE 
with $\davg{\cdots}$ an average over potentially random initial
conditions from which the dynamics is started. The matrix $D$
finally is given by $D_{tt'}=1+C_{tt'}$ for all $t,t'$ and we have 
\BE
f_t=\sum_{t'}[(\id+G)^{-1}(\rho_0\id+G')]_{tt'} \EE 
with 
\BE
G'_{tt'}=\lim_{N\to\infty}N^{-1}\sum_i \lambda_i \frac{\partial
  \overline{\davg{s_i(t)}}}{\partial h_{\lambda_i}(t')}\ . 
\EE 
These
order parameters are then to be determined as averages over
realisations of the effective processes and over the distribution of
$\lambda$ 
\BE
C_{tt'}&=&\int d\lambda R(\lambda) \avg{s_\lambda(t)s_\lambda(t')|
\lambda}_\star \\
G_{tt'}&=&\int d\lambda R(\lambda) \frac{\partial}{\partial 
h_\lambda(t')}\avg{s_\lambda(t)|\lambda}_\star \\
G'_{tt'}&=&\int d\lambda R(\lambda) \lambda \frac{\partial}{\partial
  h_\lambda(t')}\avg{s_\lambda(t)|\lambda}_\star 
\EE 
(where
$\avg{\cdots|\lambda}_\star$ is an average over realisations of the
effective process restricted to representative agents with comfort
level $\lambda$). The further analysis then assumes the existence of
an ergodic stationary state (in which correlation and response
functions depend only on time-differences, i.e. $C_{tt'}=C(t-t')$ and
similarly for $G$ and $G^{\prime}$ and in which the integrated responses
remain finite) and proceeds along the lines of \cite{HeimCool01}. We
will not report these further steps here, but only quote some of the
resulting equations describing the persistent order parameters in the
appendix. These time-independent order parameters are in the present problem given by
the persistent part $c=\lim_{T\to\infty}T^{-1}\sum_{\tau\leq T}C(\tau)$ 
of the correlation function and by the integrated responses
$\chi=\sum_\tau G(\tau)$ and $\chi'=\sum_\tau G'(\tau)$.

The two observables which we will focus on, namely the mean attendance
and the fluctuations of the attendance, can be computed from the
effective particle problem as follows. Similarly to \cite{HeimCool01}
one finds that   
\be \overline{\davg{A_t}}=f_t,
~~~~\overline{\davg{A_t
    A_{t'}}}=\frac{1}{2}[(\id+G)^{-1}D(\id+G^T)^{-1}]_{tt'}+f_t f_{t'}
\ee 
resulting in\footnote{Note here also that
  $\avg{\eta_\lambda(t)\eta_\lambda(t')}_\star=
2\overline{\davg{(A_t-\lambda)(A_{t'}-\lambda)}}$,
  reflecting the term $A(t)-\lambda_i$ in the update rules of the
  score difference of player $i$.}  
\be
\overline{\davg{A(t)^2}}-\overline{\davg{A(t)}}^2
=\frac{1}{2}[(\id+G)^{-1}D(\id+G^T)^{-1}]_{tt}.
\ee 
Temporal averages can then be expressed in terms of persistent
order parameters, and we have \be \avg{A}=\frac{\rho_0+\chi'}{1+\chi},
~~~\sigma^2=\frac{1}{2}\left[\frac{1+c}{1+\chi}+1-c\right] \ee The
former relation is exact, the second obtained within a common approximation following the lines of \cite{Book2}.
\section{Results}
One observes from the above effective agent problem that the
distribution of strategy biases $Q(\rho)$ enters only through its
first moment $\rho_0$ as claimed above. Furthermore it is
straightforward to check on the level of the effective process that
simultaneous shifts of the distributions of $\rho$ and $\lambda$ (i.e.
$\rho_i\to\rho_i+\Delta ~~\forall i$ at the same time as
$\lambda_i\to\lambda_i+\Delta$ with some $i$-independent
$\Delta\in\mathbb{R}$) do not effect the transients or stationary
states of the model, and it becomes clear that strategy biases and comfort levels are fully equivalent in the homogeneous case as already found in \cite{ChalMarsOtti03}. This is also verified in simulations. Thus we can
concentrate on the case $\rho_0=0$.

\subsection{Heterogeneous comfort levels at consistent strategies}
We here impose consistency of the strategy biases with the individual comfort levels, i.e. that  $\rho_i=\lambda_i$ for all $i$. If the $\lambda_i$
are then {\em homogeneous} over the population of agents, i.e. if
$\lambda_i=\rho_i\equiv \rho_0$ for all $i$, then the problem can be seen to
be equivalent to the standard MG with random external information
\cite{ChalMarsOtti03}. Due to the invariance under uniform shifts in the $\rho_i$ and $\lambda_i$ this equivalence of the model with homogeneous consistent strategies to the MG holds irrespectively of $\rho_0$. 

We now turn to El-Farol games with a heterogeneous distribution
of the $\lambda_i$ and $\rho_i$, still maintaining consistency $\lambda_i=\rho_i$. We here choose the simplest possible
case, namely a bi-modal distribution \be
R(\lambda)=\frac{1}{2}\left[\delta(\lambda-\varepsilon)
  +\delta(\lambda+\varepsilon)\right] \ee to study the effects of
heterogeneity in the comfort levels on the behaviour of the model,
with $\varepsilon>0$ a model parameter measuring the degree of
heterogeneity. $\varepsilon=0$ reproduces the standard MG.

\begin{figure}[t]
\vspace*{1mm}
\begin{tabular}{cc}
\epsfxsize=72mm  \epsffile{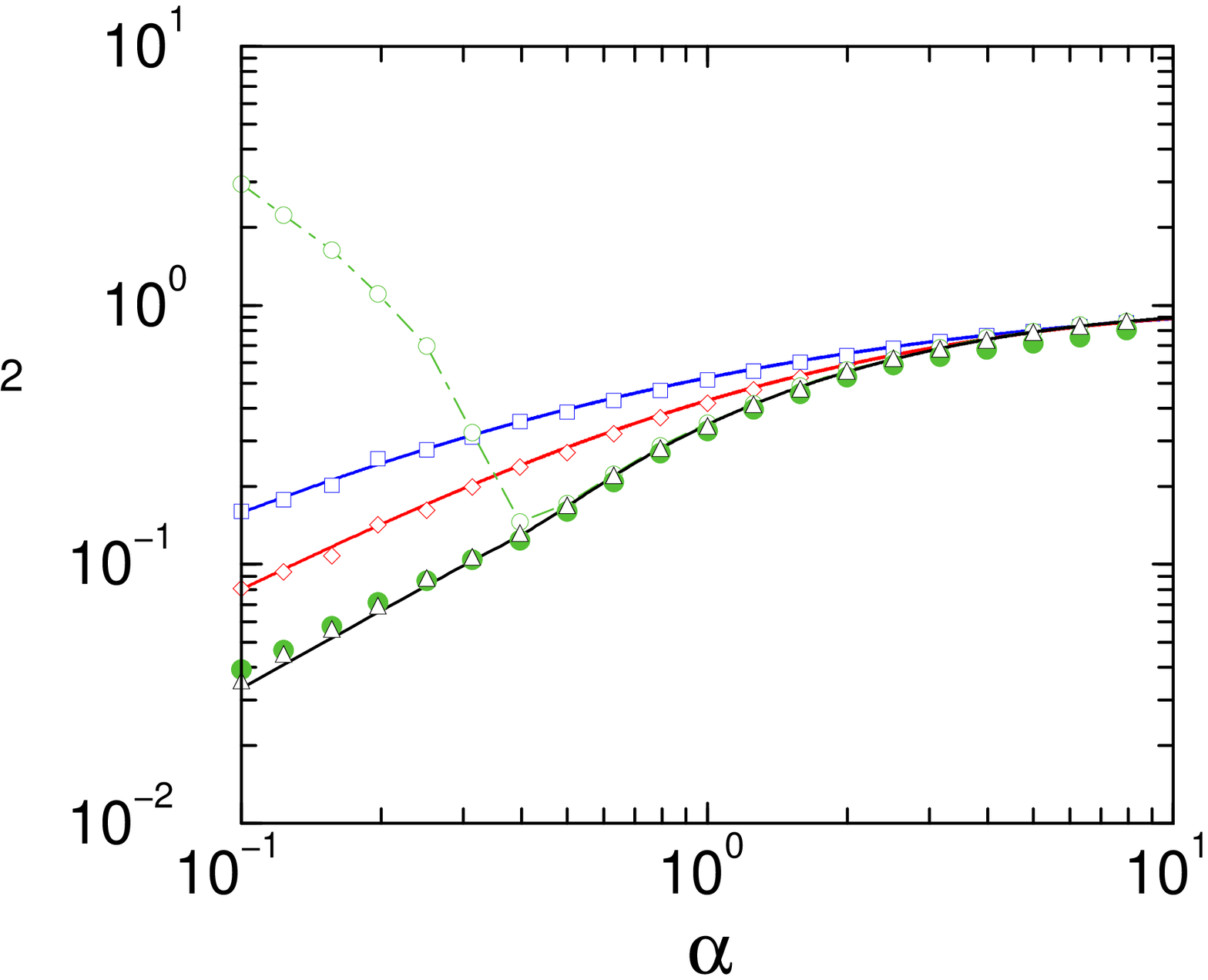} ~&~~
\epsfxsize=72mm  \epsffile{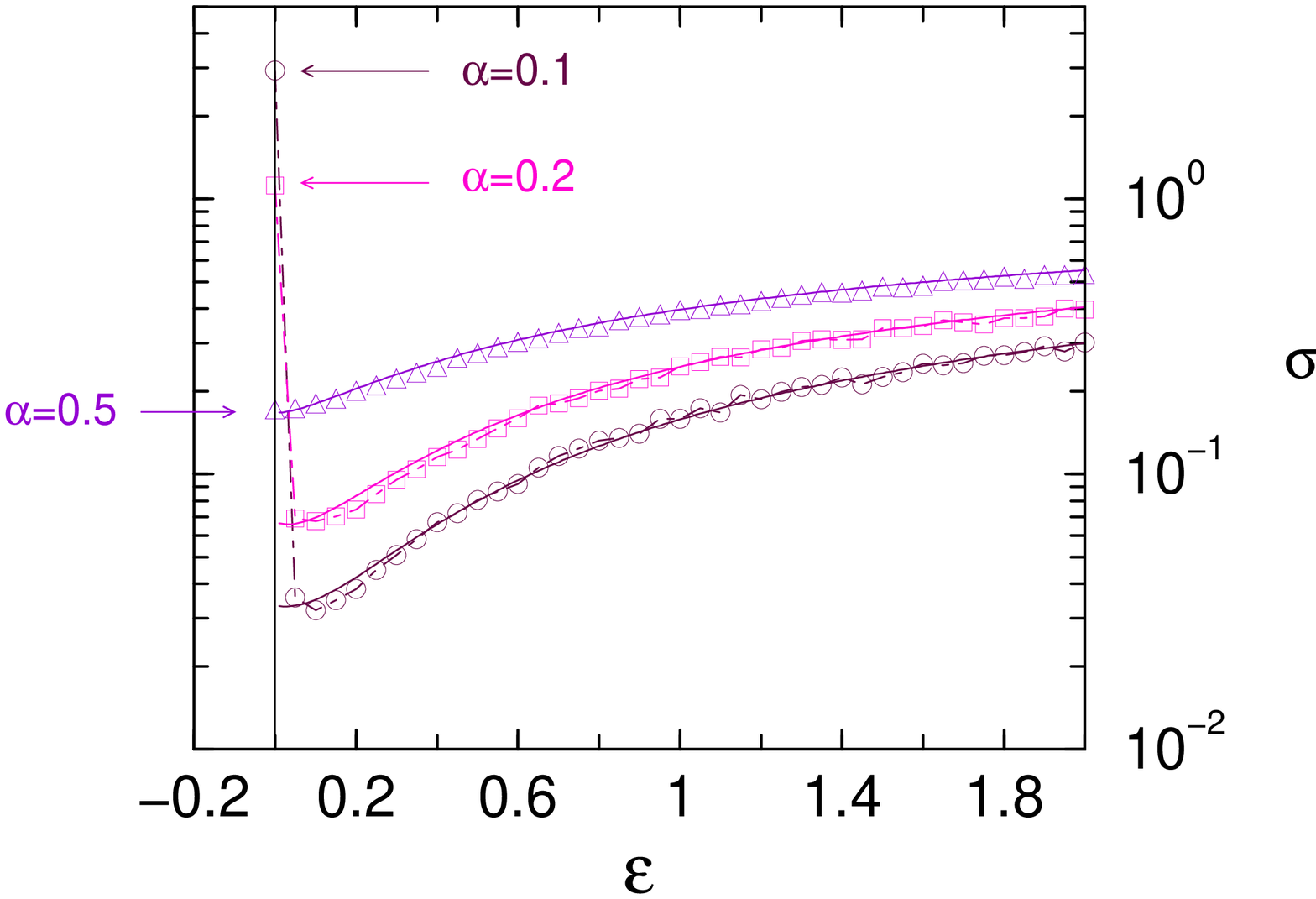}
\end{tabular}
\vspace*{4mm} \caption{(Colour on-line) Volatility for the game with consistent strategies ($\rho_i=\lambda_i~\forall i$) and bi-modal distribution
  $R(\lambda)=[\delta(\lambda+\varepsilon)+\delta(\lambda-\varepsilon)]/2$.
  Left: $\sigma^2$ as a function of $\alpha$ for different
  $\varepsilon=0$ (circles), $\varepsilon=1$ (squares),
  $\varepsilon=0.5$ (diamonds) and $\varepsilon=0.01$ (triangles).
  Open symbols are from simulations of the corresponding on-line games
  started from tabula rasa initial conditions $q_i(0)=0$ (with $N=300$
  agents, averages over $50$ samples are taken, simulations run for
  $50000$ on-line steps (or more for small $\varepsilon$)). Solid
  lines are from the analytical theory. Filled circles are for the
  standard MG ($\varepsilon=0$) from strongly biased initial
  conditions. Right: $\sigma^2$ as a function of $\varepsilon$ at
  fixed $\alpha=0.1,0.2,0.5$ (circles, squares, triangles
  respectively). Symbols are simulations of the on-line process
  started from tabula rasa initial conditions, lines the analytical
  theory in the ergodic phase. Arrows point out that there are
  discontinuities of $\sigma^2$ at $\varepsilon\to 0$ for
  $\alpha<\alpha_c$ but not at $\alpha=0.5>\alpha_c$. }
\label{fig:het}
\end{figure}

One first realises that the mean attendance $\avg{A}$ is given by the
mean comfort level $\lambda_0=\int d\lambda R(\lambda)\lambda=0$. In
the theoretical analysis this is easily seen to be the case as
$\chi'=0$ due to symmetry with respect to $\varepsilon\leftrightarrow
-\varepsilon$, and is also confirmed in simulations (not shown). The
magnitude of the fluctuations of the attendance about this level is
shown in Fig. \ref{fig:het}, both as a function of $\alpha$ at fixed
$\varepsilon$ (left) and as a function of $\varepsilon$ at fixed
$\alpha$ (right).  For $\varepsilon=0$ one finds the functional
dependence of $\sigma^2$ on $\alpha$ typical of the well-known MG,
with a minimum attained at an intermediate $\alpha_c\approx 0.33$ and
both high-volatility and low-volatility branches at $\alpha<\alpha_c$
depending on initial conditions, as shown in the left panel of
Fig. \ref{fig:het} for comparison, with open circles marking unbiased
starts ($q_i(0)=0$), and filled circles starts from strongly biased
initial conditions ($|q_i(0)|\gg 1$). One finds that $\sigma^2$
diverges as $\alpha\to 0$ for unbiased starts, and that $\sigma^2\to
0$ for strongly biases initial conditions \cite{Book2,Book3}. For
$\alpha>\alpha_c$ the starting point of the dynamics has no influence
on the persistent order parameters in the stationary state. In the
theoretical analysis the phase transition point separating both
regimes is identified as the point at which the integrated response
$\chi$ diverges, marking the breakdown of the assumptions regarding
ergodicity.

The behaviour of the model with {\em heterogeneous} (but consistent)
comfort levels is fundamentally different. For $\varepsilon>0$ all
curves $\sigma^2$ are found to be increasing functions of $\alpha$,
and the volatility remains low at low values of $\alpha$ (for all
values of $\varepsilon$ tested here), as shown in the left panel of
Fig. \ref{fig:het}. We here depict measurements from simulations
(markers) as well as the predictions of the analytical theory
(lines), and find excellent agreement. No significant dependence on
initial conditions is found. The theoretical analysis shows no sign
of a divergence of either $\chi$ or $\chi'$ so that we conclude that
the ergodicity-broken phase of the standard MG is absent as soon as
any degree of heterogeneity is added to the comfort levels. We have
also performed an analysis along the lines of \cite{HeimDeMa01} to
check for a breakdown of weak long-term memory assumptions at finite
integrated response, and could find no such a memory-onset
transition. Thus the system is fully ergodic at all $\alpha>0$ for
$\varepsilon>0$.

The marked difference between the models with and without heterogeneity in the comfort levels is demonstrated in plots of $\sigma^2$ versus $\varepsilon$ at different values of $\alpha$ in the right panel of Fig. \ref{fig:het}. For $\alpha>\alpha_c$ one observes a smooth dependence of $\sigma^2$ on $\varepsilon$ with no discontinuity at $\varepsilon\to 0$. For $\alpha<\alpha_c$ however, a characteristic jump occurs at $\varepsilon\to 0$ if simulations from zero initial conditions are considered. For all $\varepsilon\neq 0$ $\sigma^2$ remains low, whereas a substantially higher volatility is found at $\varepsilon=0$. The magnitude of the jump here depends on the choice of $\alpha<\alpha_c$ an increases as $\alpha$ becomes smaller. Note that a similar behaviour has been found in MGs with impact correction or dilution \cite{Book1,Book2,dilute}.

Finally, while we have presented results only for bimodal
distributions of the $\lambda_i$ we note that the theory presented
here straightforwardly applies to more general distributions
$R(\lambda)$ as well. Simulations of the model with Gaussian $R(\lambda)$ of
different non-zero widths $\varepsilon$ demonstrate that the phase
transition is absent also in this case and that qualitative behaviour
of the model is then similar to that of the game with a bi-modal
distribution of the $\lambda_i$. Thus the specific distribution of the
comfort levels seems to be irrelevant as far as the absence of the
phase transition is concerned, and the only relevant factor appears to
be the presence or otherwise of any type of heterogeneity.

\subsection{Inconsistent strategies}

\begin{figure}[t]
\vspace*{1mm}
\epsfxsize=80mm  ~~~~~~~\epsffile{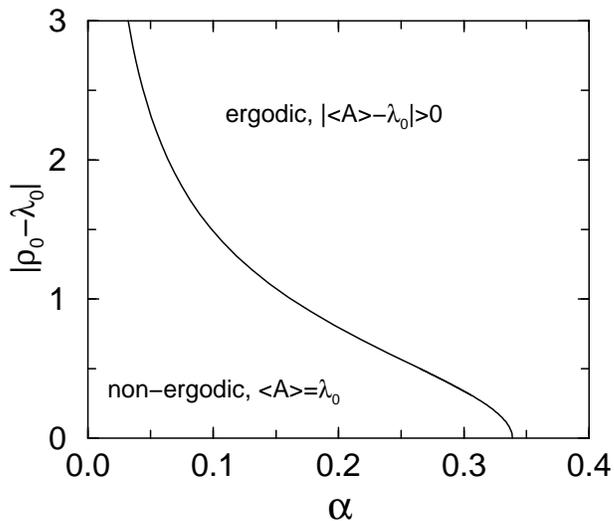} 
\vspace*{4mm} \caption{Phase diagram of the model with 
homogeneous resource level and homogeneous strategy bias ($\rho_i\equiv\rho_0$ and 
$\lambda_i\equiv\lambda_0$ for all $i$) in the 
$(\alpha,|\rho_0-\lambda_0|)$ plane. Up to re-scaling the diagram 
is identical to that 
found in \cite{ChalMarsOtti03} by static methods.}
\label{fig:phasediagram_hom}
\end{figure}

We here consider the case in which $\lambda_i\neq \rho_i$. The case of
uniformly inconsistent strategies has been studied in
\cite{ChalMarsOtti03}. In our notation the results of
\cite{ChalMarsOtti03} correspond to choosing $\rho_i\equiv\rho_0$
and $\lambda_i\equiv\lambda_0$ for all $i$, where $|\lambda_0-\rho_0|$
measures the degree of inconsistency.  The resulting phase diagram is
shown in Fig. \ref{fig:phasediagram_hom}, and corresponds to that
derived in \cite{ChalMarsOtti03} by different methods.  The dependence
only on the combination $|\lambda_0-\rho_0|$ reflects the equivalence
of strategy biases and comfort levels in the homogeneous case. Without
loss of generality, we will (mostly) consider the case $\rho_0=0$.  At
any $\lambda_0$ one then finds a phase transition of the type
which is observed in the standard MG, with an ergodic phase at
$\alpha\geq \alpha_c(\lambda_0)$, and non-ergodic behaviour below
$\alpha_c$. In the ergodic regime time-translation invariance is
maintained and the integrated response $\chi$ remains finite. The
phase transition line in Fig. \ref{fig:phasediagram_hom} is marked by
$\chi\to\infty$ (note that in the case of homogeneous strategy biases
$\chi'$ is a multiple of $\chi$ so that a divergence of $\chi'$ is
equivalent to one of $\chi$).

\begin{figure}[t]
\vspace*{1mm}
\begin{tabular}{cc}
\epsfxsize=72mm  \epsffile{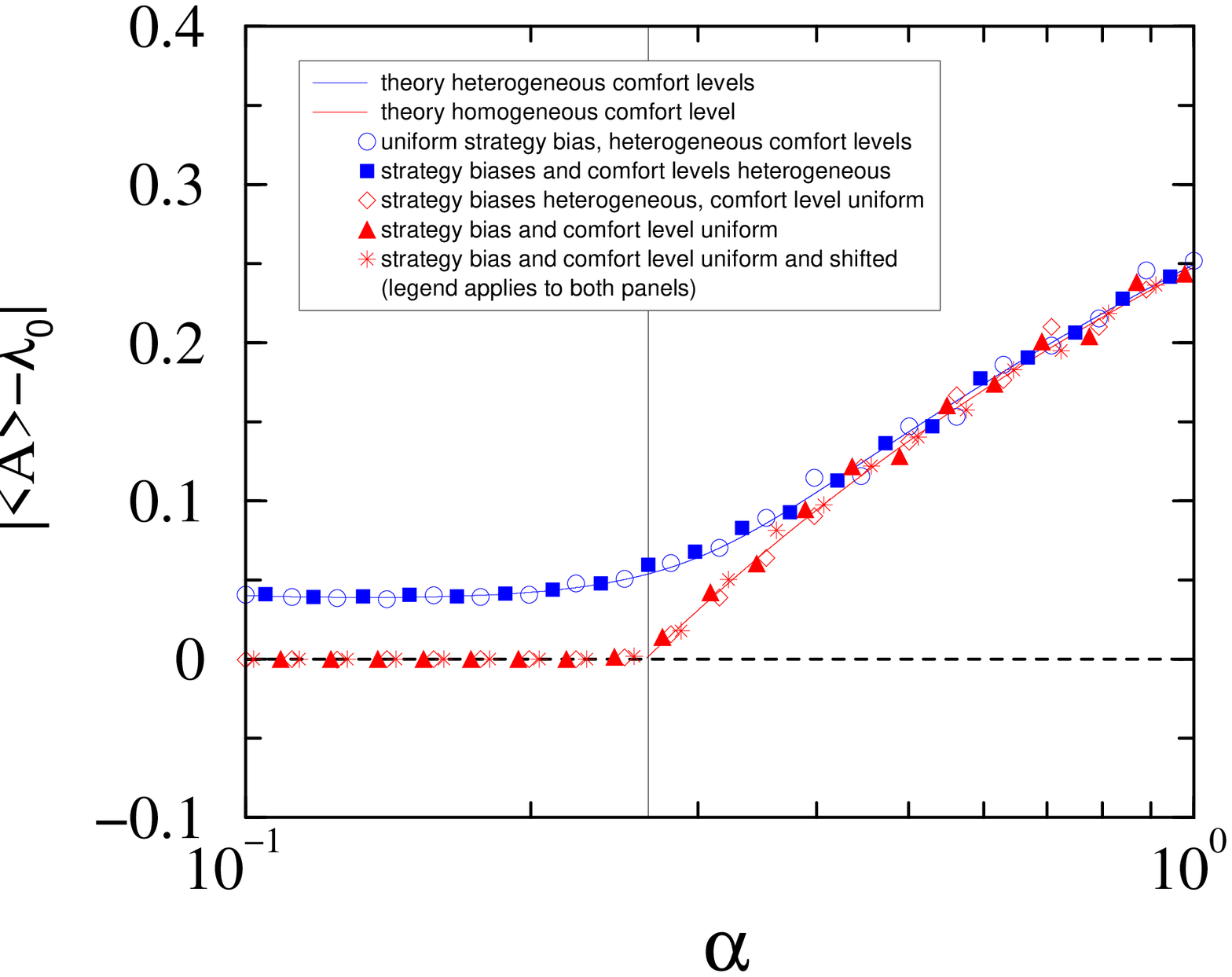} ~~&~~
\epsfxsize=72mm  \epsffile{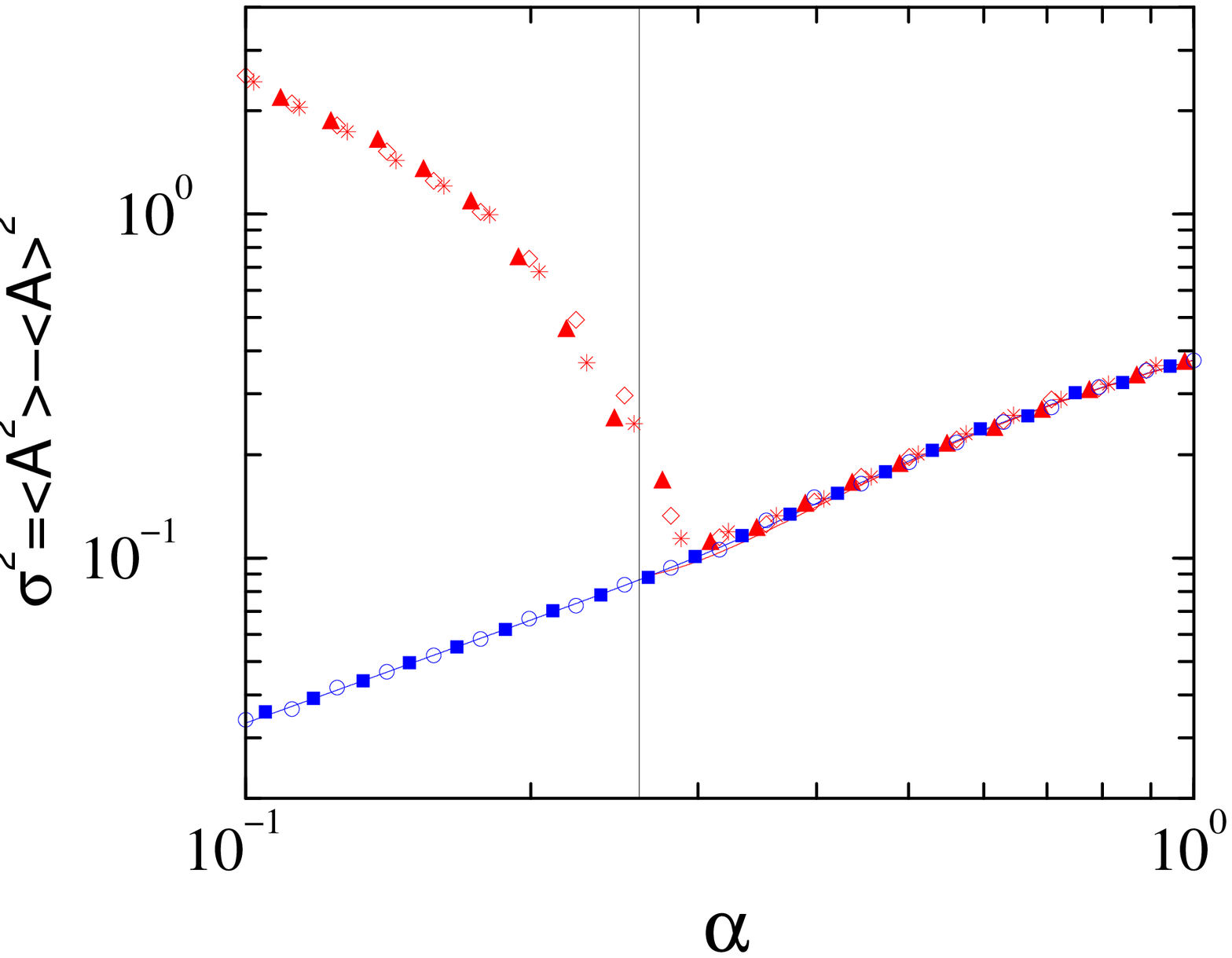}
\end{tabular}
\vspace*{4mm} \caption{(Colour on-line) Deviation of the mean
  attendance $\avg{A}$ from the mean comfort level $\lambda_0$ (left),
  and fluctuations of the attendance (right). Solid lines are from
  theory, markers from simulations, unbiased starts, $N=1000$ agents,
  run for $10^5$ steps, averaged over $10$ samples. Triangles refer to
  a system with uniform strategy bias and uniform comfort level
  ($\rho_i\equiv 0,~ \lambda_i\equiv 0.5$), stars to the shifted case
  $\rho_i\equiv -0.5, \lambda_i\equiv 0$.  Diamonds, circles and
  squares represent cases (i), (ii) and (iii) as detailed in the main
  text.  Vertical lines mark the phase transition for the model with
  uniform comfort level $\lambda_i\equiv\lambda_0=0.5$ as found from
  the analytical theory.}
\label{fig:lambdarho}
\end{figure}

It is here interesting to study the convergence of the mean attendance
$\avg{A}$ to the mean comfort level $\lambda_0$. As shown in
\cite{ChalMarsOtti03}, one finds that indeed $\avg{A}=\lambda_0$ below
$\alpha_c$, but not above (if $\lambda_0\neq 0$, for $\lambda_0=0$ one always has $\avg{A}=0$ by symmetry). In other words at fixed $\alpha$ agents are
able to recover the mean comfort level even in the inconsistent case,
provided strategy biases and comfort levels are uniform and the
inconsistency smaller than some critical value
$\lambda_{0c}(\alpha)$. This is illustrated in the left panel of
Fig. \ref{fig:lambdarho} (lower curves). At the same time the
transition between the phases in which the mean comfort level can be
retrieved, and the one where it cannot, coincides with the
ergodicity-breaking transition. At fixed $\lambda_0$ the system is
ergodic for $\alpha>\alpha_c(\lambda_0)$, and non-ergodic for
$\alpha<\alpha_c(\lambda_0)$, with large volatilities for unbiased
starts at low $\alpha$, and small volatilities for biased starts below
the transition. Measurements of the volatility for unbiased starts are
shown in Fig. \ref{fig:lambdarho} (right panel, upper curves).

We now again turn to the effects of heterogeneity on this system. In
the present case of inconsistent strategies one has to distinguish
heterogeneities in the comfort levels $\lambda_i$ from those in the strategy
biases $\rho_i$. As we will see below strategy biases and comfort
levels are no longer equivalent in the heterogeneous case, and it
makes a crucial difference whether heterogeneities are added to one or
the other. In order to disentangle the effects of both we study the
following cases:
\begin{enumerate}
\item heterogeneous strategy biases at homogeneous comfort level
  ($\rho_i=\pm 0.05$ randomly and $\lambda_i\equiv 0.5$, diamonds in Fig. \ref{fig:lambdarho})
\item homogeneous strategy biases at heterogeneous comfort levels
  ($\rho_i\equiv0$ and $\lambda_i=0.5 \pm 0.05$, circles in Fig. \ref{fig:lambdarho})
\item strategy biases and comfort levels both heterogeneous ($\rho_i=\pm
  0.05$ and $\lambda_i=0.5 \pm 0.05$ with no correlation between
  $\rho_i$ and $\lambda_i$, squares in Fig. \ref{fig:lambdarho})
\item strategy bias and comfort level both homogeneous ($\rho_i\equiv 0, \lambda_i\equiv 0.5$, triangles in Fig. \ref{fig:lambdarho}) 
\end{enumerate}
(iv) is the case of homogeneous comfort levels and strategy biases as
discussed above. One finds that the transition as observed in (iv) is
preserved only in case (i) where comfort levels are homogeneous, but
that it is absent in cases (ii) and (iii) where comfort levels are
heterogeneous. We illustrate this by plotting the mean attendance in
the stationary states for all cases in the left panel of
Fig. \ref{fig:lambdarho}. While the mean comfort level $\lambda_0=0.5$ is
successfully retrieved at low $\alpha$ in (i), systematic deviations
are present irrespectively of $\alpha$ in (ii) and (iii). The right panel of the figure demonstrates that the volatility remains low in cases (ii) and (iii) (heterogeneous comfort levels) at low $\alpha$ for unbiased starts whereas $\sigma^2$ diverges as $\alpha\to 0$ for the systems (i) and (iv) where the comfort level is uniform. Note also the good agreement of the numerical experiments with the analytical theory in the ergodic regimes, up to small deviations due to finite-size and equilibration effects.  Finally we also report simulations of a system with homogeneous comfort levels and strategy biases, but shifted by an amount $\Delta=-0.5$ with respect to (iv), i.e. we choose $\rho_i\equiv\rho_0=-0.5$ and $\lambda_i\equiv\lambda_0=0$ for all $i$ (stars in Fig. \ref{fig:lambdarho}). Results are identical to those of (iv), confirming the translation invariance against simultaneous uniform shifts of the comfort levels and strategy biases.

\subsection{Fluctuating comfort levels}
Finally, we consider comfort levels which fluctuate in time in this
section. Time-dependent comfort levels present a form of disorder
which is not quenched so that the standard analytical tools are not
straightforwardly applicable to this case. All results presented in
this section therefore rely on numerical simulations of the on-line
process (\ref{eq:scores}). Fluctuating comfort levels are here
implemented by drawing each $\lambda_i=\lambda_i(t)$ from a bimodal
distribution at every time step with no correlations in time
($\overline{\lambda_i(t)\lambda_i(t')}-\overline{\lambda_i(t)}~\overline{\lambda_i(t')}=0$). We
here distinguish between {\em collectively} and {\em individually} fluctuating comfort levels,
\BE
\mbox{collectively fluctuating:}&~~~~~&\lambda_i(t)\equiv\lambda_0+\varepsilon\zeta(t), \nonumber \\
\mbox{individually fluctuating:}&~~~~~&\lambda_i(t)\equiv\lambda_0+\varepsilon\zeta_i(t).\nonumber
\EE
Here $\zeta(t)$ and the $\zeta_i(t)$ are drawn at random at every time
step from the set $\{-1,1\}$ with equal probability, with no
correlations over time (or between players in the individually
fluctuating case). All strategy tables are generated at the beginning
of the game, and remain fixed thereafter. We limit the discussion to
the case of vanishing strategy biases ($\rho_i=0$ for all $i$) and
study models with consistent and with inconsistent
strategies. Consistency here refers to consistency as a time-average,
i.e. to $\lambda_0=\avg{\lambda_i(t)}=\rho_0=0$ for the consistent
case and to $\lambda_0\neq 0$ otherwise.

\begin{figure}[t]
\vspace*{1mm}
\begin{tabular}{cc}
\epsfxsize=72mm  \epsffile{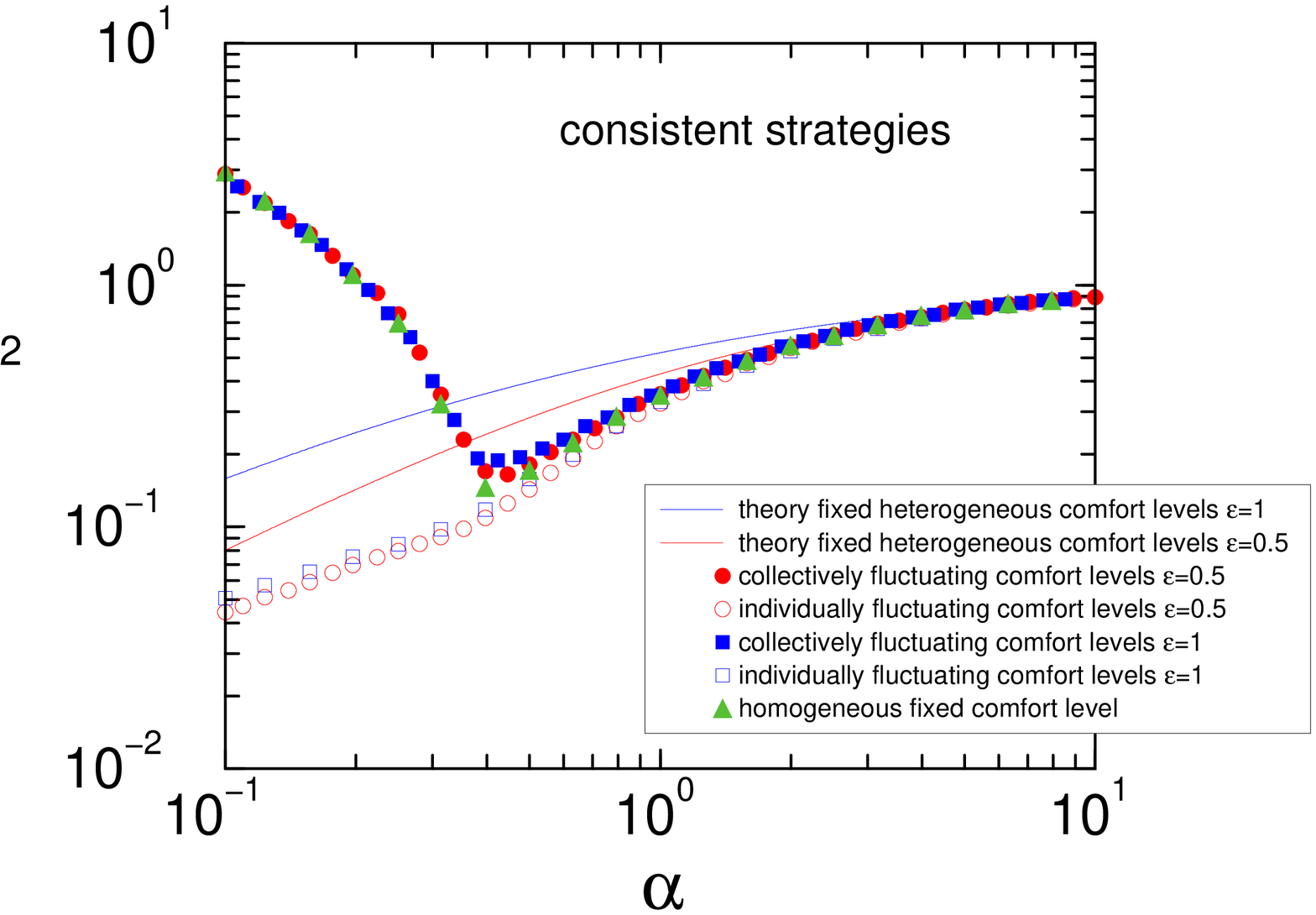} ~~&~~
\epsfxsize=72mm  \epsffile{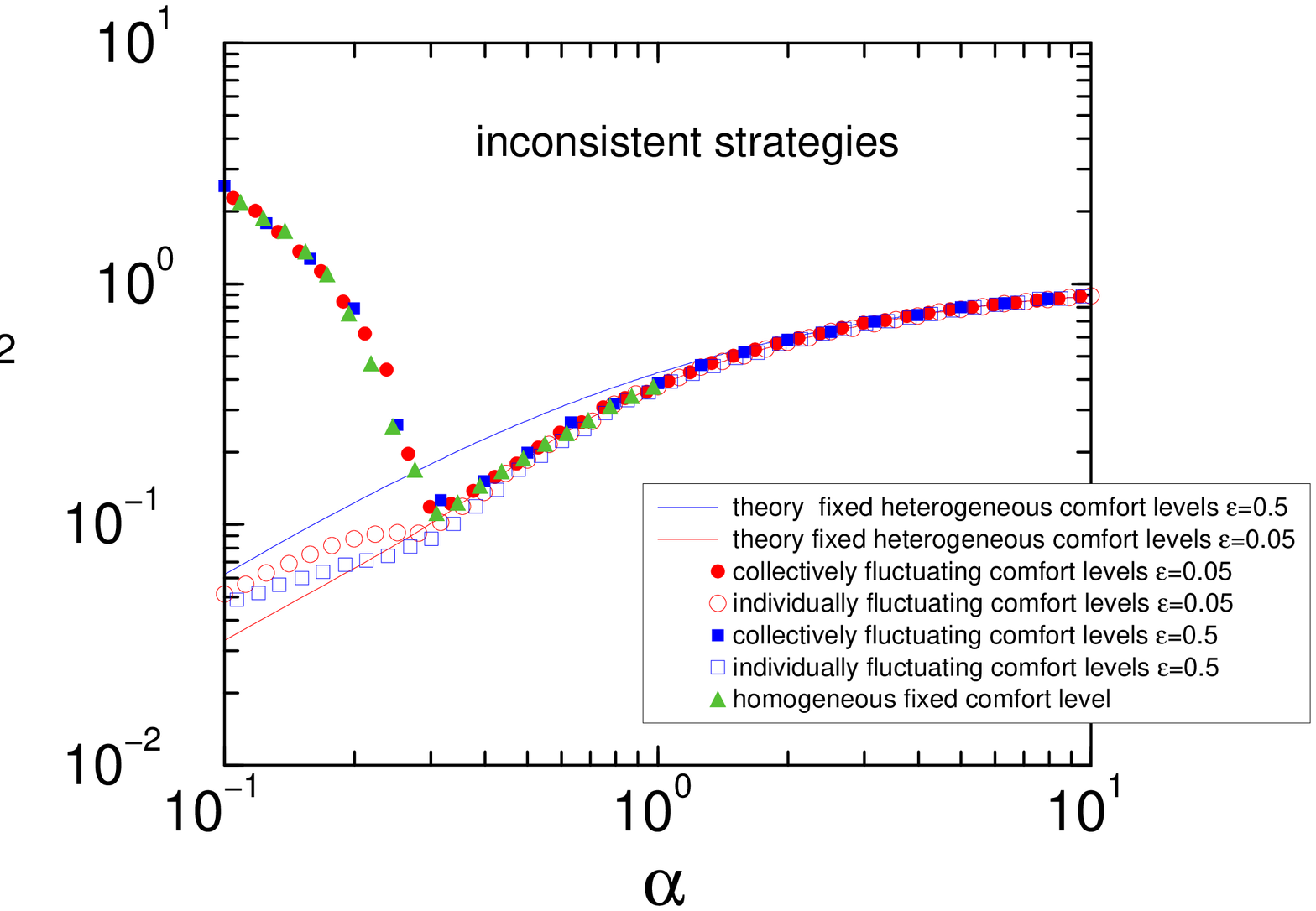}
\end{tabular}
\vspace*{4mm} \caption{(Colour on-line). Game with resource levels
  fluctuating in time. Left: strategies consistent on average, right:
  inconsistent (see text for details). Simulations are for $N=300$
  agents in the consistent case and $N=1000$ agents for inconsistent
  strategy assignments. Solid lines in both panels are for comparison
  only, and show the theoretical lines for heterogeneous comfort
  levels at same spread $\varepsilon$, but fixed in time. Filled triangles are simulations of the model with time-independent homogeneous comfort level.}
\label{fig:fluct}
\end{figure}

Results are shown in Fig. \ref{fig:fluct}. We will discuss the cases of collectively and individually fluctuating comfort levels separately in the following, and focus on the volatility of the attendance.

\subsubsection{Collectively fluctuating comfort levels:} the corresponding volatilities are indicated as solid circles and squares in Fig. \ref{fig:fluct}, with the
consistent case shown in the left panel, and the inconsistent case in
the right one. We show results for two different magnitudes
$\varepsilon$ in each case. For comparison we also display the
analytical curves for fixed comfort levels (at the same degree of
inconsistency as the simulations with fluctuating levels) as triangles
in both panels. One thus concludes that a collectively fluctuating
comfort level does not cause any noticeable effect on the resulting
volatility of the system, but that the system instead behaves as if
the collective comfort level was fixed at its time-average. 
\subsubsection{Individually fluctuating comfort levels:} the case of individually fluctuating comfort levels appears to be
crucially different from the game with collectively modulated comfort
level. Results are displayed as open markers in Fig. \ref{fig:fluct},
and one finds that individually fluctuating comfort levels reduce the
volatility significantly at small and intermediate values of $\alpha$
as compared to the system with homogeneously fixed or collectively
fluctuating levels. At larger values of $\alpha$ no such effect is
found. While we cannot fully control equilibration in our experiments,
due to apparently increased relaxation times, our simulations seem to
suggest that individually fluctuating levels reduce the volatility
below the one observed in a system with fixed heterogeneous comfort
levels of the same spread $\varepsilon$ (indicated as lines in
Fig. \ref{fig:fluct} for comparison). In particular we find that after
sufficiently long $\varepsilon$-dependent waiting times, the observed
volatility shows only little sensitivity to the numerical value of
$\varepsilon$.

\section{Geometrical interpretation of the phase transition}

Our findings regarding the absence of the phase transition in El-Farol
models with heterogeneous comfort levels is interesting also from the
point of view of statistical mechanics of the MG. In this final section we will discuss a geometrical interpretation of the phase transition of the original MG and will make some suggestions of how it may be possible to approach the problem of characterising MG models according to the presence or absence of the phase transition marked by a diverging integrated response.

\begin{figure}[t!]
\vspace*{-6mm} \hspace*{35mm} \setlength{\unitlength}{1.3mm}
\begin{picture}(120,55)
\put(-8,-0){\epsfysize=50\unitlength\epsfbox{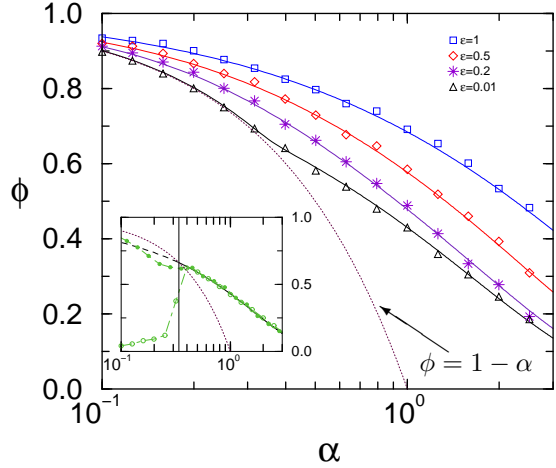}}
\put(35,12){\footnotesize $\phi=1-\alpha$}
\put(38,14){\vector(-3,2){7}}
\end{picture}
\vspace*{-0mm} \caption{(Colour on-line) Fraction of frozen agents
  $\phi$ as a function of $\alpha$. Main panel: model with consistent
  strategies, and heterogeneous fixed comfort levels ($\rho_i\equiv
  0~\forall i$ and $\lambda_i=\pm\varepsilon$ randomly with
  $\varepsilon=1,0.5,0.2,0.01$ from top to bottom. Symbols are from
  simulations of the batch process ($N=300$ agents, averages over $10$
  samples, run for a minimum of $1000$ batch steps, and longer if
  required for equilibration). Solid lines are predictions of the
  analytical theory. Inset: standard MG ($\varepsilon=0$), open
  symbols represent simulations with tabula rasa initial conditions,
  full symbols biased starts. Vertical line in the inset marks
  $\alpha_c$ where $\alpha_c=1-\phi(\alpha_c)$.}\label{fig:frozen}
\end{figure}

The update rules of most known versions of the MG are of the form
$q_i(t+1)=q_i(t)+\xi_i^{\mu(t)}A^{\mu(t)}[\bq(t)]$ (in the on-line
formulation), with some global time-dependent quantity
$A^{\mu(t)}[\bq(t)]$ which depends on the information pattern $\mu(t)$
presented to the agents at $t$ and on the strategy score differences
$\bq(t)=(q_1(t),\dots,q_N(t))$ of all agents. A geometrical picture
behind the transition of the standard MG can here be devised as
follows \cite{heimel, continuum}: due to the above update rule, the
$N$-dimensional vector $\bq(t)$ will move in the space spanned by the
$\alpha N$ vectors $\bxi^\mu=(\xi_1^\mu,\dots,\xi_N^\mu)$,
$\mu=1,\dots,\alpha N$. We will abbreviate this space by $V_{\alpha
N}(\{\bxi\})=\{\sum_{\mu=1}^{\alpha N}c_\mu
\bxi^\mu|c_1,\dots,c_{\alpha N}\in\mathbb{R}\}$ in the
following. Since a number $\phi N$ of agents can generally been shown
to `freeze' in MGs (where the fraction of frozen agents
$\phi=\phi(\alpha)$ depends on the details of the model), i.e. to
employ one strategy only and to have $|q_i(t)|\to\infty$ in the
long-time limit, the effective number of degrees of freedom is reduced
to $[1-\phi(\alpha)]N$, so that $\bq(t)$ can be thought of as having
$[1-\phi(\alpha)]N$ free components and moving in the $\alpha
N$-dimensional vector space $V_{\alpha N}(\{\bxi\})$. If $\alpha >
1-\phi(\alpha)$ then any perturbation on $\bq(t)$ can be washed out by the
dynamics. If however $\alpha <1-\phi(\alpha)$ this may not be
the case (as the dynamics is restricted to movements in $V_{\alpha
N}(\{\bxi\})$), and initial conditions may become relevant. The point at which
ergodicity breaking occurs (and the integrated response diverges) can
thus be identified as $\alpha_c=1-\phi(\alpha_c)$, which is indeed fulfilled at the transition point of the standard MG, as illustrated in the inset of Fig. \ref{fig:frozen}.

This picture breaks down whenever the update rules are not of a form
which moves the vector $\bq(t)$ in the space spanned by the
$\{\bxi^\mu\}$, specifically if $\bq(t+1)-\bq(t)$ is not a linear
combination of the $\{\bxi^\mu\}$, and one might not necessarily
expect to see a transition with diverging integrated response in this
case. Several examples can be listed here. In grand-canonical MGs
\cite{gcmg} one has update rules of the (schematic) form
$q_i(t+1)-q_i(t)=-\xi_i^{\mu(t)}A^{\mu(t)}-\kappa$, with a model
parameter $\kappa$, and the transition is absent as soon as
$\kappa\neq 0$. MGs with finite score memory \cite{mem} have update
rules of the form $q_i(t+1)-q_i(t)=-\gamma q
_i(t)-\xi_i^{\mu(t)}A^{\mu(t)}$ (with $0\leq\gamma\leq 1$ defining the
inverse time-scale over which scores are forgotten), and again the
transition seems absent as soon as $\gamma>0$. No analytical results
are available for games with finite score memory though. Finally the
transition marked by a diverging integrated response is also absent in
MGs with impact correction and with dilution, but instead preceded by
a memory-onset transition (at which $\chi$ remains finite). In both
cases one has $q_i(t+1)-q_i(t)=-\xi_i^{\mu(t)}A_i^{\mu(t)}$ where
$A_i^{\mu(t)}$ carries an explicit dependence on $i$, hence
invalidating the above picture of $\bq(t)$ moving in $V_{\alpha
N}(\{\bxi\})$. Specifically $A_i^{\mu}=A^\mu(t)-(\nu/\sqrt{N})\xi_i^\mu
s_i(t)$ in the model with impact correction (with $\nu$ measuring the
ability of the agents to correct for their own impact), and
$A_i^\mu(t)=N^{-1/2}\sum_j (c_{ij}/c)\{\omega_j^\mu+\xi_j^\mu
s_j(t)\}$ in the model with dilution (where $c_{ij}\in\{0,1\}$ and
$P(c_{ij}=1)=c$).

In the model studied in this paper model one has
$A_i^{\mu}=A^\mu-\lambda_i$ with $\lambda_i$ the comfort level of
player $i$. Thus the quantity $A_i$ does not depend on $i$ if
$\lambda_i\equiv\lambda_0$ and the above geometric interpretation
holds and one is not surprised to find the typical MG-transition. For
heterogeneous $\{\lambda_i\}$ this is no longer the case, the picture
of the vector $\bq(t)$ moving in $V_{\alpha N}(\{\bxi\})$ breaks down
and the transition is absent.

Having the listed examples in mind, one may thus speculate whether the
property of the update rules of the standard MG of restricting the
increments of $\bq(t)$ to $V_{\alpha N}(\{\bxi\})$ is a necessary
condition for a phase transition with diverging integrated response to
occur. Note that such a dynamics would probably not be a sufficient
condition for this type of transition to be present as additionally
the condition $\alpha=1-\phi(\alpha)$ has to be met at
$\alpha_c$. Interestingly we find that in our model with heterogeneous
comfort levels this condition is never fulfilled, as
$\phi(\alpha)>1-\alpha$ for all tested values of $\alpha$ (see
Fig. \ref{fig:frozen}). Only as the degree of heterogeneity
$\varepsilon$ approaches zero we do find that $\phi(\alpha)\to
1-\alpha$ for $\alpha<\alpha_c$ (where $\alpha_c$ is the transition
point of the MG with uniform comfort level, $\varepsilon=0$). Similar
observations can be made for the GCMG. In order to test our hypothesis
further, one may want to look for a model in which the dynamics is not
restricted to $V_{\alpha N}(\{\bxi\})$, but in which there is an
$\alpha$ so that $1-\phi(\alpha)=\alpha$. The MG with finite score
memory might here be a candidate as no frozen agents are present at
any $\alpha$ after long-enough equilibration \cite{mem}. Unfortunately
no analytical information on its phase behaviour is available at
present.


\section{Concluding remarks}
We have investigated a mathematical formulation of the El-Farol bar
problem focusing on the effects of heterogeneity in the comfort levels
and/or the biases of the agents' strategies. Generalizing earlier
results of \cite{ChalMarsOtti03} our main finding consists in the
observation that heterogeneity in the comfort levels even of
infinitesimal degree removes the phase transition of the standard MG,
while no such removal is observed for heterogeneous strategy
biases. This transition has also been shown to be present in an
El-Farol bar problem with {\em homogeneous} comfort level. There one
finds a phase at low but non-zero inconsistency of the strategies, in
which the mean attendance converges to the comfort level, and a second
phase in which the comfort level cannot be retrieved due to too large
an inconsistency in the strategy vectors. In the phase of successful
retrieval fluctuations of the attendance around the comfort level are
large (for tabula rasa starts), but depend on initial conditions. In
the model with heterogeneous (i.e. agent-dependent) comfort levels the
attendance converges to the mean comfort level if and only if
strategies are fully consistent.  This property is lost in the
presence of heterogeneous comfort levels for any degree of
inconsistency at any finite $\alpha$.  At the same time models with
heterogeneous comfort levels appear to show low fluctuations of the
attendance, and the typical high-volatility branches of the model with
homogeneous levels are absent. Most interestingly heterogeneous
strategy biases at uniform comfort levels do not have the same effect,
the transition is preserved. Hence we can strictly trace removal of
the transition back to the spread in comfort levels, and find that the
equivalence of comfort levels and strategy biases of the homogeneous
case does not carry over to the heterogeneous one.

We have also studied El-Farol games with temporally fluctuating
comfort levels and find that the behaviour of the volatility remains
unchanged if the level fluctuates collectively for all
players. Individually fluctuating levels, however, can reduce the
volatility significantly.

Finally we have pointed out the similarities of the present model with
other MG-type systems with and without phase transitions, and have
discussed a geometrical interpretation of the updated rules, which may
allow to characterise MG-models according to the presence or otherwise
of the ergodicity breaking transition marked by a singular integrated
response. We hope that these suggestions may stimulate further
investigation of such models with a focus on their classification
according to the types of their respective phase transitions.
\section*{Acknowledgements}
This work was supported by the European Community's Human Potential
Programme under contract HPRN-CT-2002-00319, STIPCO and by EVERGROW,
integrated project No. 1935 in the complex systems initiative of the
Future and Emerging Technologies directorate of the IST Priority, EU
Sixth Framework. TG would like to thank David 
Sherrington for fruitful discussions.


\section*{Appendix: Equations describing the ergodic stationary states}
Assuming time-translation invariance (i.e. $C_{tt'}=C(t-t')$ and
similarly for $G_{tt}$ and $G'_{tt'}$) and finite integrated response
one follows the standard ansatz to proceed from the effective agent
problem to explicit equations characterising the relevant persistent
order parameters of the ergodic stationary states. In our problem
these are given by $c$, the persistent part of the correlation
function, and by $\chi=\sum_\tau G(\tau)$ and $\chi'=\sum_{\tau}
G'(\tau)$. Further details on this analysis relying on a separation of
so-called fickle and so-called frozen agents can be found in
\cite{Book2}.

The resulting $3\times 3$ system of non-linear equations 
for $\{c,\chi,\chi'\}$ then reads
\BE
\hspace{-2cm}c=\int d\lambda ~R(\lambda) \widetilde c(\lambda), ~~~ \chi=\int 
d\lambda ~R(\lambda) ~\widetilde\chi(\lambda), ~~~ \chi'=\int d\lambda 
~R(\lambda) \lambda ~\widetilde \chi(\lambda)
\EE
with $R(\lambda)$ the distribution from which the strategy 
biases $\lambda_i$ are drawn and where $\widetilde\chi(\lambda)$ 
and $\widetilde c(\lambda)$  given by
\BE
\widetilde\chi(\lambda)&=&\frac{(1+\chi)}{\alpha}\erf
\left(\frac{\sqrt{\alpha}}{\sqrt{2g(\lambda)}(1+\chi)}\right),\\
\widetilde c(\lambda)&=&1-\erf\left(\frac{\sqrt{\alpha}}{\sqrt{2g(\lambda)}(1+
\chi)}\right) +\left[\frac{(1+\chi)^2g(\lambda)}{\alpha}\right]\nonumber \\
&&\hspace{-3cm}\times \left\{\erf\left(\frac{\sqrt{\alpha}}{\sqrt{2g(\lambda)}(1+
\chi)}\right)-\sqrt{\frac{2\alpha}{\pi g(\lambda)(1+
\chi)^2}}\exp\bigg(-\frac{\alpha}{2g(\lambda)(1+
\chi)^2}\bigg)\right\}, 
\EE
and where $g(\lambda)$ is the persistent part of the temporal 
correlations of the noise $\eta_\lambda(t)$ in the 
effective agent problem which reads
\be 
g(\lambda)=\frac{1+c}{(1+\chi)^2}+2\frac{(\rho_0+
\chi')^2}{(1+\chi)^2}-4\lambda\frac{\rho_0+
\chi'}{1+\chi}+2\lambda^2\ .
\ee

Following \cite{Book2} the fluctuations of the attendance can be approximated as
\be
\avg{A^2}-\avg{A}^2=\frac{1}{2}\frac{1+c}{(1+\chi)^2}+\frac{1}{2}(1-c),
\ee
and the mean attendance level turns out to be 
\be
\avg{A}=\frac{\rho_0+\chi'}{1+\chi}.
\ee

\end{document}